\documentclass[a4paper,11pt]{article}
\usepackage[force]{feynmp-auto}
\pdfoutput=1 

\usepackage{jheppub} 
\usepackage{pdfpages}
\usepackage[T1]{fontenc} 
\usepackage{float}
\title{\boldmath A new anomaly free flipped 341 model}

\author{Meriem Djouala,}
\author{Noureddine Mebarki and Habib Aissaoui}
\affiliation{Laboratoire de physique mathématique et subatomique.\\
	Frères Mentouri university,\\Costantine 1, Algeria}
\emailAdd{djouala.meriem@umc.edu.dz} \emailAdd{nnmebarki@yahoo.fr}
 \emailAdd{aissaoui\_h@yahoo.com}
 \abstract{
 A new flipped $SU(3)_{C}\otimes SU(4)_{L}\otimes U(1)_{X}$ model without exotic electric charges is proposed. All the quarks families are arranged in the same representation while leptons generations are in different representations leading to a tree level FCNC. Moreover, It is shown that the cancellation of the triangle chiral anomalies requires new additional leptons 10-plet and quadruplet. All fermions masses have been also discussed. Furthermore, using the most recent experimental data of the branching ratios of $\mu \longrightarrow ee\overline{e}$ and $\mu \longrightarrow e\gamma$ rare decay modes, stringent bounds on the heavy neutral bosons masses and the muon-electron  mixing matrix element are obtained.} 
\begin{document}
\maketitle \flushbottom
\section{Introduction}
\label{sec:intro}
~~Neutrinos oscillation phenomenon, dark matter, replication of quarks families, charge quantization and many other fundamental questions reveal that the Standard Model is an effective gauge field theory. Thus, going beyond the Standard Model (BSM) becomes mandatory to explain these outstanding unsolved problems.  In this paper, we are interested in a specific BSM model based on the Lie gauge group $SU(3)_{C}\otimes SU(4)_{L}\otimes U(1)_{X}$ (denoted by 341 for short) \cite{ref1}.  Many interest has been devoted to this kind of models especially those related to the LHC physics~\cite{ref1,ref7}. The most attractive feature of those models is the explanation of the family replication coming from the triangle gauge anomaly cancellation which together  with QCD asymptotic freedom, requires that only three fermion families might be present in the spectrum, as in the 331 models. Moreover, it reproduces all the SM success including the global fit to electroweak precision data. In general, requiring the gauge anomaly cancellation and avoiding the existence of fractional leptons charges and since $SU(2)_{L}\otimes U(1)_{Y}$ of the Standard Model (SM) can be embedded into
$SU(4)_{L}\otimes U(1)_{X}$ in more than one way, these models have to depend on two parameters $\beta$ and $\gamma$ \cite{ref5}. In fact, in order to assign the fermionic content of the model, it is useful to recall that the electric charge operator is written as a linear combination of the diagonal generators $T_{3}$, $T_{8}$, $T_{15}$ and X of the gauge group $SU(4)_{L}\otimes U(1)_{X}$. This allows us to write the electric charge  assignment for the fields in the fundamental representation of $SU(4)_{L}$ as:
\begin{equation}\label{eq:So}
\dfrac{Q}{e}=T_{3}+\beta T_{8}+\gamma T_{15}+X
\end{equation}
\indent It is worth to mention that the 341 models were also classified according to the scalars sectors~\cite{ref5,ref7}. Moreover, the self consistency of any BSM  model is to be free from gauge anomalies. For the case of our interest  and as it was pointed out before,  the cancellation of the $[SU(4)_{L}]^{3}$ gauge anomaly in all versions of the 341 models is fulfilled if the three quarks generations belong to different  $SU(4)_{L}$ representations: two with a left handed chirality $Q_{iL}$ (i=1,2) lie in the fundamental representation 4, whereas , the third one $Q_{3L}$
together with the three leptons generations $\psi_{lL} (l=e,\mu,\tau)$ have to transform under the conjugate fundamental representation $\bar{4}$ (or vice versa). The right handed  quarks transform as singlets. For the scalar content  with desirable properties concerning the symmetry breakdown and fermion masses including neutrinos, one can take four quartets $\phi_{k}$ (k=1,4) which are in the conjugate representation $\bar{4}$. Table \ref{tab:aa}  shows particles content in all possible 341 models parameterized by $\beta$ and $\gamma$, and the corresponding 331 sub-representations. The 321 sub-representations embedded in the fundamental and its conjugate representations of SU(3) are given by:
\begin{eqnarray}
\psi_{l}&\supset&(1,2,\frac{-1}{2})\oplus(1,1,\frac{-1}{2}-\frac{\sqrt{3}\beta}{2})\oplus(1,1,\frac{-2\gamma}{\sqrt{6}}-\frac{\beta}{2\sqrt{3}}-\frac{1}{2}),\\
Q_{i}&\supset&(3,2,\frac{1}{6})\oplus(3,1,\frac{1}{6}+\frac{\sqrt{3}\beta}{2})\oplus(3,1,\frac{1}{6}+\frac{\beta}{2\sqrt{3}}+\frac{2\gamma}{\sqrt{6}}),\\
Q_{3}&\supset&(3,2,\frac{1}{6})\oplus(3,1,\frac{1}{6}-\frac{\sqrt{3}\beta}{2})\oplus(3,1,\frac{1}{6}-\frac{\beta}{2\sqrt{3}}-\frac{2\gamma}{\sqrt{6}}),\\
\phi_{1}&\supset&(1,2,\frac{-1}{2})\oplus(1,1,\frac{-1}{2}+\frac{\sqrt{3}\beta}{2})\oplus(1,1,\frac{-\beta}{\sqrt{3}}+\frac{2\gamma}{\sqrt{6}})\\
\phi_{2}&\supset&(1,2,\frac{-\sqrt{3}\beta}{2})\oplus(1,1,0)\oplus(1,1,\frac{-1}{2}+\frac{\beta}{2\sqrt{3}}+\frac{2\gamma}{\sqrt{6}})\\
\phi_{3}&\supset&(1,2,\frac{1}{2})\oplus(1,1,\frac{1}{2}+\frac{\sqrt{3}\beta}{2})\oplus(1,1,\frac{1}{2}+\frac{\beta}{2\sqrt{3}}+\frac{2\gamma}{\sqrt{6}})\\
\phi_{4}&\supset&(1,2,\frac{1}{2})\oplus(1,1,\frac{1}{2}+\frac{\sqrt{3}\beta}{2})\oplus(1,1,0).
\end{eqnarray}
\begin{table}[h]
	\centering
	{\footnotesize 
		\begin{tabular}{|c|c|c|c|c|c|c|}
			\hline
			Name &341 representation&331 representation&Components&F\\
			\hline
			$\psi_{l}$&(1,4,$\frac{-1}{2}-\frac{\beta}{2\sqrt{3}}-\frac{\gamma}{2\sqrt{6}})$&(1,3,$\frac{-1}{2}-\frac{\beta}{2\sqrt{3}}$)$\oplus$(1,1,$\frac{-2\gamma}{\sqrt{6}}-\frac{\beta}{2\sqrt{3}}-\frac{1}{2}$)&($\nu_{l},l,E_{l}^{q_{1}},N_{l}^{q_{2}}$)&3\\
			$l^{c}$&(1,1,1)&(1,1,1)&$l^{c}$&3\\
			$l_{X}^{c}$&(1,1,$\frac{1}{2}+\frac{\sqrt{3}}{2}\beta$)&(1,1,$\frac{1}{2}+\frac{\sqrt{3}}{2}\beta$)&$l_{X}^{c}$&3\\
			$Q_{i}(i=1,2)$&(3,$\bar{4}$,$\frac{1}{6}+\frac{\beta}{2\sqrt{3}}+\frac{\gamma}{2\sqrt{6}})$&(3,$\bar{3}$,$\frac{1}{6}+\frac{\beta}{2\sqrt{3}}$)$\oplus$(1,1,$\frac{1}{6}+\frac{\beta}{2\sqrt{3}}+\frac{2\gamma}{\sqrt{6}}$)&$(u,d,D_{i}^{q_{1}},U_{i}^{q_{2}})$&2\\
			$Q_{3}$&(3,4,$\frac{1}{6}-\frac{\beta}{2\sqrt{3}}-\frac{\gamma}{2\sqrt{6}})$&(3,3,$\frac{1}{6}-\frac{\beta}{2\sqrt{3}}$)$\oplus$(1,1,$\frac{1}{6}-\frac{\beta}{2\sqrt{3}}-\frac{2\gamma}{\sqrt{6}}$)&$(d,u,U_{3}^{q_{1}},D_{3}^{q_{2}})$&1\\
			$u^{c}$&($\bar{3}$,1,$\frac{-2}{3}$)&($\bar{3}$,1,$\frac{-2}{3}$)&$u^{c}$&3\\
			$d^{c}$&($\bar{3}$,1,$\frac{1}{3}$)&($\bar{3}$,1,$\frac{1}{3}$)&$d^{c}$&3\\
			$U^{c}_{1,2}$&($\bar{3}$,1,$\frac{-1}{6}-\frac{\sqrt{3}\beta}{2}$)&($\overline{3}$,1,$\frac{-1}{6}-\frac{\sqrt{3}\beta}{2}$)&$U^{c}_{1,2}$&2\\
			$U^{c}_{3}$&($\bar{3}$,1,$\frac{-1}{6}+\frac{\sqrt{3}\beta}{2}$)&($\overline{3}$,1,$\frac{-1}{6}+\frac{\sqrt{3}\beta}{2}$)&$U^{c}_{3}$&1\\
			$D^{c}_{1,2}$&($\bar{3}$,1,$\frac{-1}{6}-\frac{\beta}{2\sqrt{3}}-\frac{2\gamma}{\sqrt{6}}$)&($\bar{3}$,1,$\frac{-1}{6}-\frac{\beta}{2\sqrt{3}}-\frac{2\gamma}{\sqrt{6}}$)&$D^{c}_{1,2}$&2\\
			$D^{c}_{3}$&($\bar{3}$,1,$\frac{-1}{6}+\frac{\beta}{2\sqrt{3}}+\frac{2\gamma}{\sqrt{6}}$)&($\bar{3}$,1,$\frac{-1}{6}+\frac{\beta}{2\sqrt{3}}+\frac{2\gamma}{\sqrt{6}}$)&$D^{c}_{1,2}$&1\\
			\hline
			$\phi_{1}$&(1,$\bar{4}$,$\frac{-\beta}{\sqrt{3}}+\frac{\gamma}{2\sqrt{6}})$&(1,$\bar{3}$,-$\frac{1}{2}+\frac{\beta}{2\sqrt{3}}$)$\oplus$(1,1,$-\frac{\beta}{\sqrt{3}}+\frac{2\gamma}{\sqrt{6}}$)&($\phi_{1}^{q_{1}}$,$\phi_{1}^{q_{2}}$,$\phi_{1}^{q_{3}}$,$\phi_{1}^{q_{4}})$&1\\
			$\phi_{2}$&(1,$\bar{4}$,$\frac{-1}{2}+\frac{\beta}{2\sqrt{3}}+\frac{\gamma}{2\sqrt{6}})$&(1,$\bar{3}$,$-\frac{\beta}{\sqrt{3}}$)$\oplus$(1,1,$-\frac{1}{2}+\frac{\beta}{2\sqrt{3}}+\frac{2\gamma}{\sqrt{6}}$)&($\phi_{2}^{q_{1}}$,$\phi_{2}^{q_{2}}$,$\phi_{2}^{q_{3}}$,$\phi_{2}^{q_{4}})$&1\\
			$\phi_{3}$&(1,$\bar{4}$,$\frac{1}{2}+\frac{\beta}{2\sqrt{3}}+\frac{\gamma}{2\sqrt{6}})$&(1,$\bar{3}$,$\frac{1}{2}+\frac{\beta}{2\sqrt{3}}$)$\oplus$(1,1,$\frac{1}{2}+\frac{\beta}{2\sqrt{3}}+\frac{2\gamma}{\sqrt{6}}$)&($\phi_{3}^{q_{1}}$,$\phi_{3}^{q_{2}}$,$\phi_{3}^{q_{3}}$,$\phi_{3}^{q_{4}})$&1\\
			$\phi_{4}$&(1,$\bar{4}$,$\frac{-3\gamma}{2\sqrt{6}})$&(1,$\bar{3}$,$\frac{1}{2}$+$\frac{\beta}{2\sqrt{3}}$)$\oplus$(1,1,0)&($\phi_{4}^{q_{1}}$,$\phi_{4}^{q_{2}}$,$\phi_{4}^{q_{3}}$,$\phi_{4}^{q_{4}})$&1\\
			\hline
	\end{tabular}}
	\caption{\label{tab:aa} Particles content of the 341 models for
		generic $\beta$ and $\gamma$ parameters where F represents the number of flavors.}
\end{table}
\indent It turns out that this quarks and leptons replication is not the only way to have a model free from the $SU(4)_{L}\otimes U(1)_{X}$  gauge anomalies. In this work and following ref \cite{ref8}, we build a new unique gauge anomaly free model without exotic electric charges baptized flipped 341 as an extension of where the previous scheme of construction is reversed that is all the quarks generations transform under the same representation while leptons are not. Thus, a flavor changing neutral current (FCNC)is expected at the tree level in the lepton sector  through the exchange of new neutral gauge bosons $Z^\prime$ and $Z^{\prime\prime}$ of the model  like in the rare leptonic decays of the form $l_{i} \longrightarrow l_{i}l_{k}\overline{l}_{k}$ and $l_{i} \longrightarrow l_{k}\gamma$ $(i\neq k)$. \\
\indent This paper is organized as follows: in section 2, we present the basic ingredients of the construction and features of the new flipped 341 model. In section 3, the anomaly cancellation is discussed. In section 4, fermions masses are emphasized at the tree level. In section 5, the FCNC and mixing in the leptonic sector are explored via the study of the rare processes $\mu \longrightarrow ee\overline{e}$ and $\mu \longrightarrow e\gamma$ and stringent inequalities and bounds on the $Z^\prime$ and $Z^{\prime\prime}$ masses are obtained. Finally, in section 6, we draw our conclusions. 
\section{A flipped 341 model}
~~In this model, the three quarks generations transform under the same conjugate fundamental representations while the leptons generations lie in different representations. It is important to mention that contrary to the ordinary 341 models, a fourth lepton generation is required in order to cancel the gauge anomaly  $[SU(4)]^{3}_{L}$ where the number of fermions
quadruplets must be equal to the number of fermions anti quadruplets. Table \ref{tab:aalo} summarizes the particles content in this model where there are no fractionally electric charged different from $\mp\frac{2}{3}$ and $\mp\frac{1}{3}$ for exotic quarks and there are no integer electric charges different from 0 and $\mp1$ for leptons and gauge bosons (model without exotic electric charges). Moreover, this model is unique, in the sense that the decomposition of the lepton 10-plet $\psi_{e}$ contains as subgroups (1,6,X)$\oplus$(1,3,X')$\oplus$(1,1,Q). To get a real (1,6,X) representation, the
value of $\beta$ should be equal to $\frac{1}{\sqrt{3}}$ \cite{ref8}. The restriction to non exotic electric charges only allows for eight different anomaly free 341 models \cite{ref12} and for a giving value of $\beta$ one has tow models with $\gamma=\frac{1}{\sqrt{6}}$ and $\gamma=\frac{-2}{\sqrt{6}}$. In the flipped version of the first model one has fractionally charged
leptons which are forbidden.\footnote{Fractionally charged leptons,
	produced abundantly in the early universe, would lead to an
	unacceptable cosmology, because there are no decay modes for the
	lightest of these exotic states \cite{ref8}.}. Therefore, the only acceptable flipped 341 is the one with  $\beta=\frac{1}{\sqrt{3}}$,
$\gamma=\frac{-2}{\sqrt{6}}$.\\
\indent The 10-plet $\psi_{e}$ and 15-plet S contain $\Sigma^{\mp}$, $\Sigma^{0}$, $\triangle^{\mp}$ and $\triangle^{0}$ which stands for
triplets sub-representations, $(\nu_{e},e)$, $(H^{+}_{S},H^{'0}_{S})$ and $(H^{''0}_{S},H^{-}_{2S})$ form $SU(2)_{L}$ doublets, ($\beta^{+}$, $\beta^{0}$, $N_{e}^{0}$, $(H^{0}_{S}$,$H^{+}_{1S}$,$H_{1S}^{'+})$ and $(H^{'0}_{1S}$,$H^{'-}_{2S}$,$H_{S}^{'-})$ compose the triplet denoted by
(1,3,$\frac{1}{3}$), (1,$\overline{3}$,$\frac{2}{3}$) and (1,3,$\frac{-2}{3}$) respectively, while,
$E_{e}^{-}$, $\sigma^{+}$, $\sigma_{S}^{0}$ and $H^{0}_{1S}$ are singlets.\\
\indent In our flipped 341 model and in order to generate masses for particles one has to have four scalar fields $\phi_{1}$, $\phi_{2}$, $\phi_{3}$, $\phi_{4}$ and a new scalar 15-plet S which transforms under the adjoint representation 15 of the group SU(4).
 	\begin{table}[H]
 		\centering
 		{\footnotesize
 			\begin{tabular}{|c|c|c|c|c|c|c||c|}
 				\hline
 				Name & 341 representation& 331 representation & components & F\\
 				\hline
 				$\psi_{e}$&(1,10,0)&(1,6,$\frac{-1}{3}$)+(1,3,$\frac{1}{3}$)+(1,1,1)&$\left(
 				\begin{array}{cccc}
 				\Sigma^{+} & \frac{\Sigma^{0}}{\sqrt{2}} & \frac{\nu_{e}}{\sqrt{2}}& \frac{\beta^{+}}{\sqrt{2}}  \\
 				\frac{\Sigma^{0}}{\sqrt{2}} &  \Sigma^{-} & \frac{e^{-}}{\sqrt{2}}& \frac{\beta^{0}}{\sqrt{2}} \\
 				\frac{\nu_{e}}{\sqrt{2}} &\frac{e^{-}}{\sqrt{2}}&E_{e}^{-}& \frac{N_{e}^{0}}{\sqrt{2}}  \\
 				\frac{\beta^{+}}{\sqrt{2}} &   \frac{\beta^{0}}{\sqrt{2}} &\frac{N_{e}^{0}}{\sqrt{2}} &\sigma^{+}\\
 				\end{array}
 				\right)$&1\\
 				$\psi_{\alpha}(\alpha=\mu,\tau)$&(1,4,$\frac{-1}{2}$)&(1,3,$\frac{-2}{3}$)+(1,1,0)&($\nu_{\alpha},l_{\alpha},E^{-}_{\alpha},N_{\alpha}^{0}$)&2\\
 				$\widetilde{\psi}$&(1,$\overline{4}$,$\frac{-1}{2}$)&(1,3,$\frac{-1}{3}$)+(1,1,-1)&($\widetilde{e}^{-},\widetilde{\nu},\widetilde{N}^{0},\widetilde{E}^{-}$)&1\\
 				$l_{\alpha}^{c}$&(1,1,1)&(1,1,1)&$l^{c}_{\alpha}$&6\\
 				$Q_{\alpha}$&(3,$\overline{4}$,$\frac{1}{6})$&(3,$\bar{3}$,$\frac{1}{3}$)+(3,1,$\frac{-1}{3}$)&$(d_{\alpha},u_{\alpha},U_{\alpha},D_{\alpha})$&3\\
 				$u^{c}$&($\overline{3}$,1,$\frac{-2}{3}$)&($\overline{3}$,1,$\frac{-2}{3}$)&$u^{c}$&6\\
 				$d^{c}$&($\overline{3}$,1,$\frac{1}{3}$)&($\overline{3}$,1,$\frac{1}{3}$)&$d^{c}$&6\\
 				\hline
 				$\phi_{1}$&(1,$\overline{4}$,$\frac{-1}{2})$&(1,$\bar{3}$,
 				$\frac{-1}{3}$)+(1,1,-1)&($\phi_{1}^{-}$,$\phi_{1}^{0}$,$\phi_{1}^{'0}$,$\phi_{1}^{'-})$&1\\
 				$\phi_{2}$&(1,$\overline{4}$,$\frac{-1}{2})$&(1,$\bar{3}$,$\frac{-1}{3}$)+(1,1,-1)&($\phi_{2}^{-}$,$\phi_{2}^{0}$,$\phi_{2}^{'0}$,$\phi_{2}^{'-})$&1\\
 				$\phi_{3}$&(1,$\overline{4}$,$\frac{1}{2})$&(1,$\bar{3}$,$\frac{2}{3}$)+(1,1,0)&($\phi_{3}^{0}$,$\phi_{3}^{+}$,$\phi_{3}^{'+}$,$\phi_{3}^{'0})$&1\\
 				$\phi_{4}$&(1,$\overline{4}$,$\frac{1}{2})$&(1,$\bar{3}$,$\frac{2}{3}$)+(1,1,0)&($\phi_{4}^{0}$,$\phi_{4}^{+}$,$\phi_{4}^{'+}$,$\phi_{4}^{'0})$&1\\
 					S&(1,15,0)&(1,8,0)+(1,3,$\frac{-2}{3}$)+(1,$\bar{3}$,$\frac{2}{3}$)+(1,1,0)&
 				$\left(
 				\begin{array}{cccc}
 				 \triangle^{+} &\frac{\triangle^{0}}{\sqrt{2}} & \frac{H^{+}_{S}}{\sqrt{2}} & \frac{H^{0}_{S}}{\sqrt{2}} \\
 				  \frac{  \triangle^{0}}{\sqrt{2}}& \triangle^{-}&\frac{H'^{0}_{S}}{\sqrt{2}} &  \frac{H^{+}_{1S}}{\sqrt{2}} \\
 				\frac{H^{''0}_{S}}{\sqrt{2}}  &  \frac{H^{-}_{2S}}{\sqrt{2}} & H_{1S}^{0} & \frac{H_{1S}^{'+}}{\sqrt{2}} \\
 				\frac{H^{'0}_{1S}}{\sqrt{2}} & \frac{H_{2S}^{'-}}{\sqrt{2}} &  \frac{H^{'-}_{S}}{\sqrt{2}} & \sigma_{S}^{0} \\
 				\end{array}
 				\right)$&1   \\
 				\hline
 			\end{tabular}
 			\caption{\label{tab:aalo} The complete anomaly free fermions content and scalar sectors in the flipped
 				341 model with their flavors (F), both $(1,3,0)+(1,2,\frac{-1}{2})+(1,1,-1)$ and $(1,3,0)+(1,2,\frac{-1}{2})+(1,2,\frac{1}{2})+(1,1,0)$ are the decomposition of $(1,6,\frac{-1}{3})$ \cite{ref8} and $(1,8,0)$ respectively.}}
 	\end{table}
 \section{Anomalies cancellation}
~~ Based on to the fact that $\mathrm{Tr}[T^{a}]=0$ and $\mathrm{Tr}[\tau^{a}]=0$ where $T^{a}$ and $\tau^{a}$ are the generators of the Lie gauge groups SU(3) and SU(4) respectively, all the triangle gauge anomalies are automatically canceled expect those of the non-trivial ones: $[SU(4)_{L}]^{3}$,
 $[SU(3)_{C}]^{2}\otimes U(1)_{X}$, $[SU(4)_{L}]^{2}\otimes U(1)_{X}$,
 $[SU(3)_{C}]^{3}$, $[U(1)_{X}]^{3}$ and $[Grav]^{2}\otimes U(1)_{X}$.
 From the particles content shown in table \ref{tab:aalo}, we notice that our flipped 341 model is free from all the gauge triangle anomalies. In fact, the total contribution of the $[SU(4)_{L}]^{3}$ anomaly comes from
 	\cite{ref99}:
 	\begin{equation}
 	\mathcal{A}^{abc}(4_{L})\bigg(\sum_{Q_{mL},f_{iL}}4_{L}-\sum_{Q_{nL}}\bar{4}_{L}\bigg)=\mathcal{A}^{abc}(4_{L})(n_{4L}-n_{\bar{4}_{L}})
 	\end{equation}
 	where the anomaly coefficient  $\mathcal{A}^{abc}(4_{L})=\mathrm{Tr}(T^{a}_{L},\{T^{b}_{L},T^{c}_{L}\})$
  with $\mathcal{A}^{abc}(4_{L})=-\mathcal{A}^{abc}(\bar{4}_{L})$ and 
 	$4_{L}$ respectively $\bar{4}_{L}$ are SU(4) quadruplet and 
 	anti-quadruplets fundamental representations. Here $n_{4L}$ and $n_{\bar{4}_{L}}$ are the number of left-handed
 	fermions quadruplets and anti-quadruplets respectively. This
 	anomaly cancels only if the number of the quadruplets in the fundamental representation $4_{L}$ equals to the number of the quadruplets in the conjugate fundamental representation $\bar{4}_{L}$ as it is the case in our model (see Table \ref{tab:aalo}). We remind that the 10-plet $\psi_{e}$ contributes as much as eight quadruplets in the group SU(4) \cite{ref9} $(\mathcal{A}(10)=8\mathcal{A}(4))$, together with the remaining lepton generations $\psi_{\mu}$ and $\psi_{\tau}$, the contribution of all lepton generations equals to  $10\mathcal{A}(4)$. While, the arrangement of the three quarks families in the fundamental conjugate representation
 	$\overline{4}$ makes their contribution equal to $9\mathcal{A}(4)$. Thus, to ensure the cancellation of the $[SU(4)_{L}]^{3}$ anomaly, a new exotic lepton $\widetilde{\psi}$ lies in the conjugate representation $\overline{4}$ must be introduced.\\
  \indent The cancellation of the $(SU(3)_{C})^{3}$ anomaly requires the introduction of the charge conjugate of each quark field as an $SU(4)_{L}\otimes U(1)_{X}$ singlet for which the quantum number X coincides with the electric charge Q \cite{ref12}.\\
  \indent The flipped 341 model is free from the $SU(4)_{L}\otimes U(1)_{X}$, $SU(3)_{C}\otimes U(1)_{X}$, $(Grav)^{2} \otimes U(1)_{X}$ and $ (U(1)_{X})^{3}$ anomalies only if its particles content satisfied the following conditions respectively:
 \begin{eqnarray}
  \frac{1}{2}\bigg(\sum X_{l}^{L}&+&3\sum X_{q}^{L}\bigg)=0,\label{eq:Q4}\\      
 \frac{1}{2}\bigg(4\sum X_{q}^{L}&-&\sum_{Sing} X_{q}^{R}\bigg)=0,\label{eq:Q3}\\
  4\sum X_{l}^{L}+12\sum X_{q}^{L}&-&3\sum X_{q}^{R}-\sum X^{R}_{l}=0,\label{eq:Q2}\\
  4\sum (X_{l}^{L})^{3}+12\sum(X_{q}^{L})^{3}&-&3\sum_{Sing}(X^{R}_{q})^{3}-\sum_{Sing}(X^{R}_{l})^{3}=0.\label{eq:Q}
  \end{eqnarray}	 
 where $X_{l(q)}^{L}$ and $X_{l(q)}^{R}$ are the quantum numbers associated
 to the $U(1)_{X}$ group of the left $(L)$ and right $(R)$ handed leptons $(l)$ and quarks $(q)$. The factors 4 appear to take into account all the components of the  quadruplets and anti-quadruplets \cite{ref99}, the factors 3 appear because the quarks are in $SU(3)_{C}$ triplets \cite{ref99} and the factors 12 represent the multiplication of $3\times 4$, the quark generations are quadruplets and triplets under the group SU(4) and SU(3) respectively.\\
 \indent The special content of the flipped 341 model ensures the cancellation of the  $(Grav)^{2}\otimes U(1)_{X}$ anomaly which requires that the sum of all the $U(1)_{X}$ charges yields to zero as it mentioned in equation \eqref{eq:Q}.
 Table \ref{tab:aaap} shows the cancellation of the gauge
 anomalies in the flipped 341 model. Notice that the sum of all rows of each column in the table \ref{tab:aaap} multiplied by
 the number of flavors (F) vanishes. Therefore, the model is gauge
 anomalies free.
\begin{table}[H]
	{\footnotesize
 		\begin{tabular}{|c|c|c|c|c|c|c|c|}
 			\hline
 			Field &$[SU(3)_{C}]^{3}$&$[SU(4)_{L}]^{3}$&$[SU(4)_{L}]^{2} U(1)_{X}$&$[SU(3)_{C}]^{2} U(1)_{X}$ &$[U(1)_{X}]^{3}$&$(Grav)^{2} U(1)_{X}$& F\\
 			\hline
 			$\psi_{e}$&0&4&0&0&0&0&1\\
 			$\psi_{\alpha}$&0&$\frac{1}{2}$&$\frac{-1}{4}$&0&$\frac{-1}{2}$&-2&2\\
 			$\widetilde{\psi}$&0&$\frac{-1}{2}$&$\frac{-1}{4}$&0&$\frac{-1}{2}$&-2&1\\
 			$l_{\alpha}^{c}$&0&0&0&0&1&1&6\\
 			$Q_{\alpha}$&2&$\frac{-3}{2}$&$\frac{1}{4}$&$\frac{1}{3}$&$\frac{1}{18}$&2&3\\
 			$u^{c}_{\alpha}$&$\frac{-1}{2}$&0&0&$\frac{-1}{3}$&$\frac{-8}{9}$&-2&6\\
 			$d^{c}_{\alpha}$&$\frac{-1}{2}$&0&0&$\frac{1}{6}$&$\frac{1}{9}$&1&6\\
 			\hline
 		\end{tabular}
 		\caption{\label{tab:aaap} Gauge anomalies fields contributions in the flipped
 			341 model.}}
 \end{table}
\section{Fermions masses}
~~~Following ref \cite{ref8}, we assume that there exists a
stable, charge-preserving vacuum state. In this spirit, we allow all neutral scalar quadruplets components in Table \ref{tab:aalo} to have a non-zero vacuum expectation value (VEV) \cite{ref8}:
$$
\langle\phi_{1}\rangle= \left(
\begin{array}{ccc}
0\\
k_{1}\\
n_{1}\\
0
\end{array}
\right)\langle\phi_{2}\rangle= \left(
\begin{array}{ccc}
0\\
k_{2}\\
n_{2}\\
0
\end{array}
\right)\langle\phi_{3}\rangle=
\left(
\begin{array}{ccc}
k_{3}\\
0\\
0\\
n_{3}
\end{array}
\right)\langle\phi_{4}\rangle=
\left(
\begin{array}{ccc}
k_{4}\\
0\\
0\\
n_{4}
\end{array}
\right)
$$
~~~Note that in order to eliminate the unwanted Yukawa Lagrangian interactions (nonphysical terms) such as terms like $\nu_{e}e$, $\Sigma^{0}e$ and many others, only the elements $S_{33}$ and $S_{44}$ develop vacuum expectation values (VEVs). Therefore, the VEV of the scalar 15-plet S is:
$$\langle S\rangle=
\left(
\begin{array}{cccc}
0&0&0&0\\
0&0&0&0\\
0&0&n_{1S}&0\\
0&0&0&n_{2S}
\end{array}
\right)
$$
The gauge symmetry in the flipped 341 model is broken to the Standard Model one via the following steps:
\begin{equation}
\begin{alignedat}{1}
	SU(3)_{c}\otimes S&U(4)_{L}\otimes U(1)_{X}\\
	&\Downarrow n_{3}, n_{4}, n_{1S}, n_{2S}\\
	SU(3)_{c}\otimes S&U(3)_{L}\otimes U(1)_{X'}\\
	&\Downarrow n_{1}, n_{2}\\
	SU(3)_{c}\otimes S&U(2)_{L}\otimes U(1)_{Y}\\
	&\Downarrow k_{1}, k_{2}, k_{3}, k_{4}\\
    SU(3)_c&\otimes U_{em}
    \end{alignedat}
\end{equation}
Where $k_{i}(i=1,2,3,4)\ll n_{1}, n_{2}\ll n_{i}(i=3,4,1S,2S)$.
Regarding the Yukawa Lagrangian $\mathcal{L}_{Y}$ it has the form:
\begin{equation}
\mathcal{L}_{Y}=\mathcal{L}_{Leptons}+\mathcal{L}_{Quarks}
\end{equation}
The lepton sector allows the combination of $SU(4)_L$ quadruples and
anti quadruplets as:
\begin{eqnarray}\label{eq:zlmp}
\mathcal{L}_{Leptons}&=&\sum_{i=1}^{2}y_{\alpha\beta}^{l(i)}\psi_{\alpha}l^{c}_{\beta}\phi_{i}+h.c
\end{eqnarray}
and for quarks:
\begin{equation}\label{eq:z}
\mathcal{L}_{Quarks}=\sum_{i=1}^{2}y_{\alpha\beta}^{u(i)}Q_{\alpha}u^{c}_{\beta}\phi^{*}_{i}+\sum_{i=3}^{4}y^{d(i)}_{\alpha\beta}Q_{\alpha}d^{c}_{\beta}\phi^{*}_{i}+h.c
\end{equation}
 where $\beta$ and $\alpha$ stand for the flavors of the fermions fields.
Notice that an additional term in the Lagrangian \eqref{eq:zlmp} is introduced in order to generate a mass to the exotic leptons $\Sigma$:
 \begin{equation}\label{eq:domi}
\mathcal{L}_{Leptons}=\frac{y'}{\Lambda}\psi_{e}^{aa'}(\psi_{e})^{bb'}(S)^{cc'}S^{dd'}\epsilon_{abcd}\epsilon_{a'b'c'd'},
 \end{equation}
Where ${y'}$ is the Yukawa coupling, $\Lambda$ represents
the cutoff scale where new physics is expected and $\epsilon_{abcd}$ is the Levi Civita tensor.\\
\indent Regarding the masses of quarks, substituting the VEVs into the Yukawa Lagrangian \eqref{eq:z}, we extract the tree level masses of the up and down quarks respectively:
$$
m^{u}= \left(
\begin{array}{cccc}
y_{\alpha\beta}^{u(1)}k_{1}+ y_{\alpha\beta}^{u(2)}k_{2}\\
y_{\alpha\beta}^{u(1)}n_{1}+ y_{\alpha\beta}^{u(2)}n_{2}\\
\end{array}
\right)
$$
$$
m^{d}= \left(
\begin{array}{cccc}
y_{\alpha\beta}^{d(3)}k_{3}+y_{\alpha\beta}^{d(4)}k_{4}\\
y_{\alpha\beta}^{d(3)}n_{3}+y_{\alpha\beta}^{d(4)}n_{4}\\
\end{array}
\right)
$$
where we have written $m^{u}$ (resp. $m^{d}$) in the basis ($u_{\alpha},U_{\alpha}$)
and ($u_{\beta}^{c}$) (resp.($d_{\alpha},D_{\alpha}$) and ($d_{\beta}^{c}$)), $y_{\alpha\beta}^{u(i)}$(i=1,2), $y_{\alpha\beta}^{d(j)}$ (j=2,3) represent the Yukawa couplings.\\
\indent At the tree level, the light charged lepton mass matrix $m^{l}$ has been extracted from the Yukawa Lagrangian \eqref{eq:zlmp} and it is written in the basis ($l_{\alpha}, E_{\alpha}$) and ($l^{c}_{\beta}$))
$$
m^{l}= \left(
\begin{array}{ccccccccccc}
y_{\alpha\beta}^{l(1)}k_{1}+ y_{\alpha\beta}^{l(2)}k_{2}\\
y_{\alpha\beta}^{l(1)}n_{1}+ y_{\alpha\beta}^{l(2)}n_{2}\\
\end{array}
\right)
$$
The mass matrix $m^{l}$ shows that the charged leptons  $\mu$ and $\tau$
acquire their masses at the
tree level except the electron, it remains massless. Moreover,
notice also that in contrast with the flipped 331 model \cite{ref8}, the following Lagrangian:
\begin{equation}
\mathcal{L}=y_{\alpha}^{l}\psi_{e}l^{c}_{\alpha}S
\end{equation}
is not allowed since it is not gauge invariant under both $U(1)_{X}$ and $SU(4)_{L}$ symmetries. Therefore, the exotic lepton $E_{e}$ and the electron e remain massless. Thus, in order to generate masses to both $E_{e}$ and e, one has to introduce  the following effective dimension-5 operator:
\begin{equation}\label{eq:mzp12}
\mathcal{L}_{eff}=\frac{\lambda^{(ij)}_{\alpha l}}{\Lambda}(\psi_{e})^{aa'}l^{c}_{\alpha}(\phi_{i})_{b}(\phi_{j})_{b'}\delta_{ab}\delta_{a'b'}.
\end{equation}
In this case:
\begin{eqnarray}
m_{E_{e}}&=&\frac{\lambda^{(ij)}_{E_{e}}}{\Lambda}n_{i}n_{j},\\
m_{e}&=&\frac{\lambda^{(ij)}_{e}}{\sqrt{2}\Lambda}(k_{i}n_{j}+n_{i}k_{j}).
\end{eqnarray}
The VEVs $n_{i}, n_{j}$ (i,j= 1,2), $\Lambda$ and $k_{i}$ are of $\mathcal{O}$ (TeV) and $\mathcal{O}$ (GeV) respectively. Taking $n_{j}$=$\Lambda$, then the electron and $E_{e}$ masses are proportional to the SM symmetry breaking $k_{i}$ and heavy scales respectively.\\
\indent Concerning the exotic leptons $\Sigma^{\mp}$ and $\Sigma^{0}$, their masses can be obtained from the Lagrangian \eqref{eq:domi} to get:
\begin{eqnarray}
m_{\Sigma^{\mp}}&=&\frac{4y'}{\Lambda} n_{1S}n_{2S},\nonumber\\ m_{\Sigma^{0}}&=&\frac{-2y'}{\Lambda} n_{1S}n_{2S}
\end{eqnarray}
Where the VEVs $n_{iS}$ (i=1,2) are of $\mathcal{O}$ (TeV). In the case of our interest the exotic leptons $\beta^{+}$, $\beta^{0}$, $N_{e}^{0}$ and $\sigma^{+}$ receive their masses from the following Lagrangian:
  \begin{equation}\label{eq:vova}
\mathcal{L}_{eff}=Y^{L}(\psi_{e})^{aa'}(\widetilde{\psi})_{b}(\phi_{j})_c \delta_{ab}\delta_{a'c}.
\end{equation}
where j=3,4. Notice that we have identified the charged and neutral elements of
$\widetilde{\psi}$ namely, $\widetilde{e}$, $\widetilde{\nu}$, $\widetilde{N}^{0}$ and $\widetilde{E}^{-}$ with the charge conjugated right-handed components of
leptons already introduced in the electron generation $\psi_{e}$ which are $\beta^{+}$, $\beta^{0}$, $N_{e}^{0}$ and $\sigma^{+}$ respectively.
$$
\widetilde{\psi}_{L}=\left(
\begin{array}{ccc}
\widetilde{e}_{L}^{-} \\
\widetilde{\nu}_{L} \\
\widetilde{N}^{0}_{L}\\
\widetilde{E}^{-}_{L} \\
\end{array}
\right)\equiv
 \left(
\begin{array}{ccc}
(\beta^{+}_{R})^{c} \\
(\beta^{0}_{R})^{c}\\
(N^{0}_{eR})^{c}\\
(\sigma^{+}_{R})^{c} \\
\end{array}
\right)
$$
We have used this identification to avoid the
presence of charged exotic lepton with masses of the order
of the electroweak scale \cite{ref99}. Therefore, the exotic leptons $\beta^{+}$, $\beta^{0}$, $N_{e}^{0}$ and $\sigma^{+}$ receive their masses denoted by $m_{L}$ and $m_{\sigma^{+}}$ respectively
from the Lagrangian \eqref{eq:vova} to get:
\begin{eqnarray}
m_{L}&=&Y^{L}\frac{n_{j}}{\sqrt{2}},\nonumber\\
m_{\sigma^{+}}&=&Y^{\sigma^{+}}n_{j}.
\end{eqnarray}
With $Y^{L}$ and $Y^{\sigma^{+}}$ are the Yukawa coupling and L stands for the exotic leptons $\beta^{+}$, $\beta^{0}$ and $N_{e}^{0}$.
\section{FCNC}
\begin{figure}[H]
	\centering 
	\includegraphics[width=.46\textwidth]{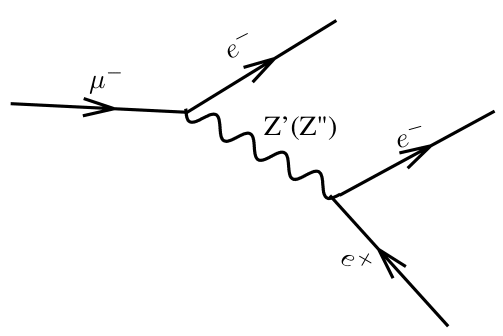}
	\hfill
	\includegraphics[width=.46\textwidth]{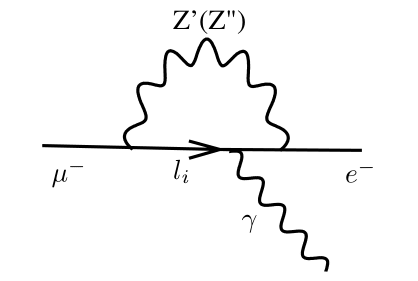}
	\caption{\label{fig:leev2} The decays $\mu \longrightarrow ee\overline{e}$ and $\mu \longrightarrow e\gamma$ via the neutral gauge bosons $Z^\prime$($Z^{\prime\prime}$).}
\end{figure}

\begin{figure}[H]
\centering 
\includegraphics[width=.46\textwidth]{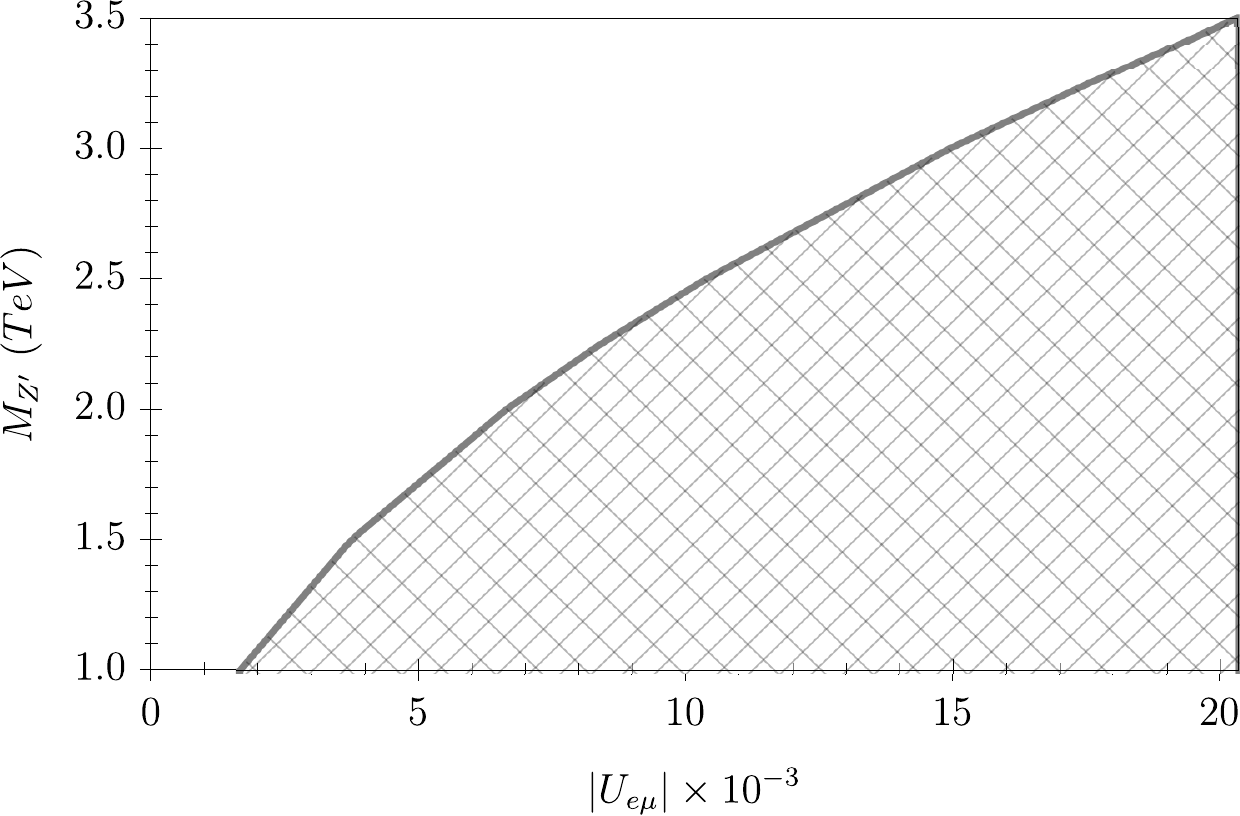}
\hfill
\includegraphics[width=.46\textwidth]{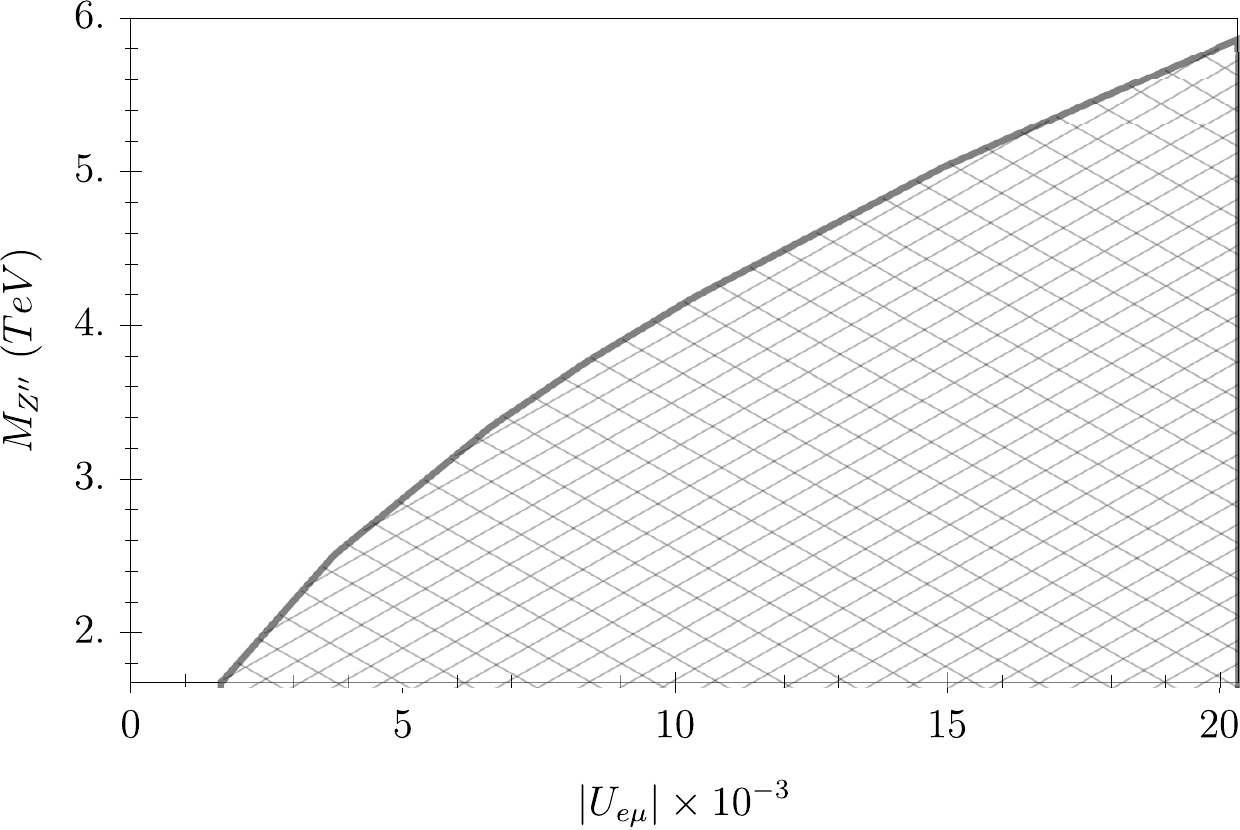}
\caption{\label{fig:eee} The variation of $M_{Z^\prime}$ and $M_{Z^{\prime\prime}}$ as a function of  $|U_{e\mu}|$.}
\end{figure}

 ~~~~~The ordinary versions of the 341 models predict the existence of new heavy neutral gauge bosons $Z^\prime$ and $Z^{\prime\prime}$ which have universal couplings with leptons, while, the quarks couplings with the new gauge bosons are non-universal. The existence of a non diagonal matrix when we rotate the flavor basis into the mass eigenstates leading to the occurrence of FCNC in the quark sector. We suppose that the major sources contributing to the lepton flavor violating processes come from the direct gauge interactions between the charged leptons and new massive gauge bosons $Z^{'}$ and $Z^{''}$, while the others such as scalar contributions are considered to be small and neglected \cite{ref13}.\\
 \indent This scheme is reversed in the flipped 341 model, since two of leptons families are arranged differently from the remaining generations whereas the quarks families lie in the same representation. Therefore, the FCNC occurs only in the lepton sector at the tree level through the exchange of the gauge bosons $Z^\prime$ and $Z^{\prime\prime}$. In the context of our model, we study the lepton flavor violation processes namely 
 $\mu \longrightarrow ee\overline{e}$ and $\mu \longrightarrow e\gamma$. In fact,
the lepton neutral currents of $Z^\prime$ and $Z^{\prime\prime}$ are described by the following Lagrangian:
\begin{eqnarray}
\mathcal{L}_{NC}\supset-g\overline{F}\gamma^{\mu}\bigg(T_{3\mu}A_{3}+T_{8}A_{8\mu}+T_{15}A_{15\mu}+Xg_{X}B_{\mu}\bigg)F
\end{eqnarray}
where $T_{i}=\lambda_{i}/ 2$ and $\lambda_{i}$ (i=3, 8, 15) are the diagonal Gell-Mann matrices in the group SU(4), g and $g_{X}$ represent the gauge couplings of the $SU(4)_{L}$ and $U(1)_{X}$ respectively, X is the charge associated to the group $U(1)_{X}$ and F runs over all fermions multiplets. Here $T_{i}\psi_{\alpha}=\frac{1}{2}\lambda_{i}\psi_{\alpha}$ (i=3, 8, 15) and  $T_{i}\psi_{e}=\frac{1}{2}(\lambda_{i}\psi_{e}+\psi_{e}\lambda_{i})$(i=3, 8, 15) \cite{ref13}, 
the right-handed leptons $e_{aR}$ and $E_{aR}$ do not
participate in FCNC \cite{ref13}. Then we obtain:
\begin{eqnarray}
\mathcal{L}_{NC}&\supset&\frac{g}{2}\bigg(\frac{1-t_{W}^{2}}{\sqrt{3-t_{W}^{2}}}\bigg)\bigg(\overline{\nu}_{L}\gamma^{\mu}T_{\nu}\nu_{L}+\overline{l}_{L}\gamma^{\mu}T_{l}l_{L}+\overline{E}_{L}\gamma^{\mu}T_{E}E_{L}+\overline{N}_{L}\gamma^{\mu}T_{\nu}N_{L}\bigg) Z^\prime_{\mu}\nonumber\\&+&\frac{g}{2}\bigg(\frac{-2}{\sqrt{6+4t^{2}_{X}}}\bigg)\bigg(\overline{\nu}_{L}\gamma^{\mu}T'_{\nu}\nu_{L}+\overline{l}_{L}\gamma^{\mu}T_{l}l'_{L}+\overline{E}_{L}\gamma^{\mu}T'_{E}E_{L}+\overline{N}_{L}\gamma^{\mu}T'_{\nu}N_{L}\bigg) Z^{\prime\prime}_{\mu}.
\end{eqnarray}
Where $t_{X}^{2}=s_{W}^{2}/(1-2s_{W}^{2})$ \cite{ref133}, $s_{W}$, $c_{W}$ and $t_{W}$ are the sine, cosine and tangent of the electroweak mixing angle. Whereas $T_{i}$ and $T'_{i}$ are:
\begin{eqnarray}
 T_{\nu}=T_{l}&=&\text{diag}\bigg(1,(-1-t_{W}^{2})/(1-t_{W}^{2}),(-1-t_{W}^{2})/(1-t_{W}^{2})\bigg),\\ T_{E}&=&\text{ diag}\bigg((4-2t_{W}^{2})/(1-t_{W}^{2}),(2-2t_{W}^{2})/(1-t_{W}^{2}),(2-2t_{W}^{2})/(1-t_{W}^{2})\bigg),\\
 T_{N}&=&\text{diag}\bigg(2/(1-t_{W}^{2}),0,0\bigg), \\
T'_{\nu}&=&T'_{l}=T'_{E}=\text{diag}\bigg(1,(1-2t_{X}^{2})/2,(1-2t_{X}^{2})/2)\bigg)\\ \text{and}\\ T'_{N}&=&\text{diag}\bigg(-1,-(3+2t_{X}^{2})/2,-(3+2t_{X}^{2})/2\bigg). 
\end{eqnarray}
 \indent Changing from the flavor basis into the mass basis $l_{L}=V_{lL}l'_{L}$, we get the following Lagrangian:
\begin{equation}
\mathcal{L}_{NC}\supset\frac{g}{2}\bigg(\frac{1-t_{W}^{2}}{\sqrt{3-t_{W}^{2}}}\bigg)\overline{l'}_{L}\gamma^{\mu}\bigg(V^{\dagger}_{lL}T_{l}V_{lL}\bigg)l'_{L} Z^\prime_{\mu}-\frac{g}{2}\bigg(\frac{-2}{\sqrt{6+4t^{2}_{X}}}\bigg)\overline{l'}_{L}\gamma^{\mu}\bigg(V^{\dagger}_{lL}T'_{l}V_{lL}\bigg)l'_{L} Z^{\prime\prime}_{\mu}.
\end{equation}
Thus,
\begin{equation}
\mathcal{L}_{NC}\supset\frac{g}{2}\bigg(\frac{1-t_{W}^{2}}{\sqrt{3-t_{W}^{2}}}\bigg)\overline{l'}_{L}\gamma^{\mu}(V^{*}_{lL})_{\alpha i}(V_{lL})_{\beta j}l'_{L} Z^\prime_{\mu}+\frac{g}{2}\bigg(\frac{-2}{\sqrt{6+4t^{2}_{X}}}\bigg)\overline{l'}_{L}\gamma^{\mu}(V^{*}_{lL})_{\alpha i}(V_{lL})_{\beta j}l'_{L} Z^{\prime\prime}_{\mu}.
\end{equation}
Here $l'$ can be $e, \nu, E, N$ and $i\neq j$ for flavor changing.\\
 \indent Regarding the decays  $\mu \longrightarrow ee\overline{e}$ and $\mu \longrightarrow e \gamma$ are among the rare decay  processes which are used to search for the charged lepton flavor violation . The corresponding Feynman diagrams are shown in Figure \ref{fig:leev2}. The first vertex in the left diagram shows the tree level FCNC coupling of $Z^\prime$($Z^{\prime\prime}$) boson (cLF changing), whereas the right diagram represents the one loop level process where $l_{i}$ can be any lepton.
Here we will consider the internal fermions line to be either $\mu$ or e, so that we will have only one FCNC $Z^\prime\mu e$ ($Z^{\prime\prime}\mu e$) vertex.
The transferred momentum, whose maximal
value is about the muon mass, is much smaller than $M_{Z^\prime(Z^{\prime\prime})}$ \cite{ref13},
therefore, the Branching ratio of the processes $\mu \longrightarrow ee\overline{e}$ and $\mu \longrightarrow e\gamma$ through the exchange of $Z^\prime$ and $Z^{\prime\prime}$ are found to be respectively: 
\begin{eqnarray}
Br(\mu \longrightarrow ee\overline{e})&=&M^{4}_{W}|U_{e\mu}|^{2}\bigg(\frac{(1-t_{W}^{2})^{2}}{
2M_{Z^\prime}^{2}(3-t_{W}^{2})}+\frac{2}{M_{Z^{\prime\prime}}^{2}(6+4t^{2}_{X})}\bigg)^{2},\\
Br(\mu \longrightarrow e\gamma)&=&\frac{24\alpha}{16\pi}M^{4}_{W}|U_{e\mu}|^{2}\bigg(\frac{(1-t_{W}^{2})^{2}}{
M_{Z^\prime}^{2}(3-t_{W}^{2})}
+\frac{4}{M^{2}_{Z^{\prime\prime}}(6+4t^{2}_{X})}\bigg)^{2}
\end{eqnarray}
where $U_{e\mu}$ is the mixing matrix for the lepton. We have neglected the contribution coming from the electron internal line as it is proportional to  $m_{e}/m_{\mu}$ \cite{ref16} and considered the electrons as massless particles.\\
 \indent Using the experimental upper limit of the branching ratio $\text{Br}(\mu \longrightarrow ee\overline{e})\leq 10^{-12}$  \cite{ref15} together with the current experimental limit $\text{Br}(\mu \longrightarrow e\gamma)\leq 10^{-13}$ \cite{ref14}, we obtain, an upper limit on the lepton flavor violating matrix element
$|U_{\mu e}|\leq 1.66\times 10^{-3} \frac{M^{2}_{Z^\prime}}{TeV^2}$ and a stringent bound on the gauge bosons masses $M_{Z^\prime}\leq 0.597 M_{Z^{\prime\prime}}$.
Now if $M_{Z^\prime}$$\approx$ $\mathcal{O}$(1 TeV), one has $|U_{\mu e}|$$\approx$$1.66 \times 10^{-3}$ and $M_{Z^{\prime\prime}}$$\approx$$\mathcal{O}$(1.68 TeV).
Figure \ref{fig:eee} shows the variation of $M_{Z^\prime}$ and $M_{Z^{\prime\prime}}$ as a function of $|U_{\mu e}|$ and the dashed areas represent the allowed region where the constraints are verified.
\section{Conclusion}
~~~~~~~In this work, we have constructed a new unique model based on the gauge
group $SU(3)\otimes SU(4)\otimes U(1)$ called the flipped 341 model
where all the three quarks families arrange in the same
representation whereas leptons generations are not. The anomalies
cancellation requires the introduction of new extra exotic leptons a 10 plet $\psi_{e}$ and a quadruplet $\widetilde{\psi}$.\\
\indent  In the flipped 341 model, a dangerous flavor changing neutral current can appear in the lepton sector through the interactions with the neutral gauge bosons $Z^\prime$ and $Z^{\prime\prime}$, whereas, it is absent in the quark sector which makes this model unable to explain the anomalies of both kaon and B decays \cite{ref8} in contrast to the ordinary versions of the 341 models. The comparison of our results with the experimental data concerning the branching ratios of the two processes  $\mu \longrightarrow ee\overline{e}$ and $\mu \longrightarrow e\gamma$ lead to a stringent constraint on the matrix element and on the masses of the gauge bosons.
 \section*{Acknowledgments}
 We  would like to thank Prof. Renato M. Fonseca for fruitful discussions.  We also thank Uladzimir Khasianevich for useful and clarifying comments. We are very grateful to the Algerian ministry of higher education and scientific research and DGRSDT for the financial support. 
 
\end{document}